\documentclass[conference]{IEEEtran}
\IEEEoverridecommandlockouts
\usepackage{cite}
\usepackage{amsmath,amssymb,amsfonts}
\usepackage{graphicx}
\usepackage{textcomp}
\usepackage{xcolor}
\usepackage{booktabs}
\usepackage[T1]{fontenc}
\usepackage{enumitem}
\usepackage[hyphens]{url}
\usepackage{hyperref}  
\usepackage{cleveref}

\usepackage{algorithm}
\usepackage{algpseudocode} 
\usepackage{authblk}
\usepackage{amssymb}
\usepackage{pifont}
\newcommand{\xmark}{\ding{55}}%
\def\BibTeX{{\rm B\kern-.05em{\sc i\kern-.025em b}\kern-.08em
    T\kern-.1667em\lower.7ex\hbox{E}\kern-.125emX}}
\usepackage{xspace}

\newcommand{\hleak}{HammerLeak\xspace}
\newcommand{\dsteal}{DeepSteal\xspace}
\newcommand{\strain}{Mean Clustering\xspace}

\usepackage{amsmath,amsfonts,bm}

\def\eqref#1{equation~\ref{#1}}

\def\1{\bm{1}}

\def\vx{{\bm{x}}}
\def\vy{{\bm{y}}}

\DeclareMathAlphabet{\mathsfit}{\encodingdefault}{\sfdefault}{m}{sl}
\SetMathAlphabet{\mathsfit}{bold}{\encodingdefault}{\sfdefault}{bx}{n}
\newcommand{\tens}[1]{\bm{\mathsfit{#1}}}

\def\tW{{\tens{W}}}

\DeclareMathOperator*{\argmax}{arg\,max}

\usepackage{multirow}
\usepackage{soul}
\usepackage{subfig}
\usepackage[scaled=.8]{beramono}

\renewcommand{\baselinestretch}{0.98}

\makeatother
\begin{document}

\title{\huge{\dsteal: Advanced Model Extractions Leveraging Efficient Weight Stealing in Memories}}
\author[1]{Adnan Siraj Rakin\textbf{*}}
\author[2]{Md Hafizul Islam Chowdhuryy\textbf{*}} 
\author[2]{Fan Yao}
\author[1]{Deliang Fan}
\affil[ ]{{\textit{\textbf{*} Co-First Authors with Equal Contributions}}}
\affil[1]{Department of Electrical,Computer and Energy Engineering, Arizona State University}
\affil[2]{Department of Electrical and Computer Engineering, University of Central Florida}

\renewcommand\Authands{ and }

\maketitle

\begin{abstract}
Recent advancements of Deep Neural Networks (DNNs) have seen widespread deployment in multiple security-sensitive domains. The need of resource-intensive training and use of valuable domain-specific training data have made these models a top intellectual property (IP) for model owners. 
One of the major threats to the DNN privacy is model extraction attacks where adversaries attempt to steal sensitive information in DNN models. Recent studies  show hardware-based side channel attacks can reveal internal knowledge about DNN models (e.g., model architectures)
However, to date, existing attacks cannot extract detailed model parameters (e.g., weights/biases). In this work, for the first time, we propose an advanced model extraction attack framework \emph{\dsteal} that effectively steals DNN weights with the aid of memory side-channel attack. Our proposed \dsteal comprises two key stages. Firstly, we develop a new weight bit information extraction method, called \emph{\hleak}, 
through adopting the rowhammer based hardware fault technique as the information leakage vector. \emph{\hleak} leverages several novel system-level techniques tailed for DNN applications to enable fast and efficient weight stealing. 
Secondly, we propose a novel substitute model training algorithm with \emph{\strain} weight penalty, which leverages the partial leaked bit information effectively and generates a substitute prototype of the target victim model. We evaluate this substitute model extraction method on three popular image datasets (e.g., CIFAR-10/100/GTSRB) and four DNN architectures (e.g., ResNet-18/34/Wide-ResNet/VGG-11). The extracted substitute model has successfully achieved more than \emph{90 \%} test accuracy on deep residual networks for the CIFAR-10 dataset. Moreover, our extracted substitute model could also generate effective adversarial input samples to fool the victim model. Notably, it achieves similar performance (i.e., $\sim$ 1-2 \% test accuracy under attack) as white-box adversarial input attack (e.g., PGD/Trades). 
\end{abstract}


\section{Introduction}

The recent development of deep learning technologies has made them an integral part of our daily life. This widespread application of Deep Neural Networks (DNNs) includes but is not limited to image classification~\cite{he2016deep}, image detection~\cite{He_2015_ICCV} and speech recognition~\cite{xiong2016achieving}, many of which are deployed in security-sensitive settings~\cite{247638}. 
DNN models typically take a tremendous amount of resources to train, and in many cases the training relies on the use of valuable domain-specific data. As a result, DNN models are regarded as the \emph{top intellectual properties} (IP) for machine learning (ML) service providers and model owners~\cite{nas}. With the rapid development of system and hardware level attack vectors that can compromise and tamper computing systems~\cite{plundervolt,rambleed,gruss2016rowhammer,kim2014flipping,voltpwn}, understanding and investigating the security of deep learning systems has become imperative. 




Model extraction attacks aim to infer or steal critical information from DNN models to achieve certain malicious goals~\cite{jagielski2020high}.
Particularly, active learning is a popular approach in recovering the performance of a victim DNN model~\cite{chandrasekaran2020exploring,orekondy2019knockoff,barbalau2020black,nayak2019zero,correia2018copycat,pal2019framework,papernot2017practical}. 
These methods primarily use input and output scores to recover the victim model's performance. It is challenging to extract the exact internal decision boundary only with input-output scores, which is especially the case for DNNs with complex and deep structures~\cite{milli2019model,jagielski2020high,rolnick2020reverse}. These techniques typically come with tremendous training overhead and substantial attack costs because of heavy model queries. 



Recent advances in hardware-based exploitation have showed that adversaries can use side channel attacks to gain sensitive information in a target system~\cite{naghibijouybari2018rendered,liu2015last,numacovert,2020branchspec}.
Particularly, several works have shown that DNN model configuration parameters and structure information can be extracted by leveraging microarchitecture attacks~\cite{yan2020cache}, physical side channels (e.g., through observing power consumption or EM emanations~\cite{batina2018csi}), and bus snooping attacks~\cite{hu2020deepsniffer} with very high accuracy. Hardware-based attacks can be extremely dangerous as they allow adversaries to \emph{directly gain internal knowledge} about the victim DNN models. However, most existing hardware-based DNN attacks focus on inferring high-level model information (i.e., commonly model architectures)\footnote{Recently, ~\cite{zhu2021hermes} shows an attack that recovers a whole model by monitoring the plaintext packets transmitted over the PCIe-bus. Such attack is essentially an \emph{overt channel} that can be mitigated via traffic encryption.}. 
It remains uncertain whether accurate information about model weights---the core information of DNN models, can be effectively exfiltrated via hardware-based side channel exploits. Note that acquisition of such information can potentially further extend DNN adversarial capabilities, including constructing substitute models with high accuracy, reproducing model fidelity (i.e., identical prediction as to the victim model), and bolster transfer adversarial attacks~\cite{papernot2017practical}. 
In this paper, we aim to answer the following question: \emph{Is it possible for an adversary to perform advanced model extractions through stealing weight parameters using hardware-based attacks?}

While obtaining model weights can be useful intuitively, there are a few major challenges from the attacker's perspective to efficiently capture and effectively utilize such information. \emph{First}, although fine-grained secret leakage has been widely shown to be plausible in many non-ML applications through hardware-based side channels (e.g., microarchitecture attacks in particular~\cite{yarom2014flush+,liu2015last, evtyushkin2018branchscope, yan2019attack,branchspectre2021}). To date, no effective side channels have been demonstrated to exfiltrate detailed model weights due to the lack of distinguishable control and data-flow dependencies in DNN applications. Second, DNN models are often extremely large (with millions of parameters), even with a hardware-based attack vector that can recover certain model weight information, it is typically impractical to assume that \emph{the entire weights} can be exfiltrated in practical settings. Moreover, prior works \cite{Rakin_2019_ICCV,yao2020deephammer} have proved that variations on only tens, out of millions, weight parameters will completely malfunction a DNN model. In this case, whether partial information about model weights can be effectively leveraged to build a stronger model extraction attack is uncertain. The final challenge involves how to design highly optimized ML techniques based on the obtained unique and partial model weight knowledge for different attack objectives. 

In this paper, we present \emph{\dsteal}, an advanced model extraction attack framework using efficient model weight stealing with the aid of hardware-based side channels. The objective of our attack is to recover (partial) weight parameters of a target DNN model (i.e., victim model), which will be harnessed to build useful substitute models using novel learning schemes. 
At a high level, \dsteal consists of two key stages. To address the aforementioned first challenge, \textbf{in the first stage}, we develop a novel hardware-based side channels that can exfiltrate partial bit information of model weight parameters, called \emph{\hleak}. Particularly, we leverage the well-known rowhammer fault attack~\cite{kim2014flipping} as the information leakage vector. Our exploitation is motivated by prior studies showing that the occurrence of rowhammer-induced fault in a memory cell highly depends on the data pattern of its neighbouring bits~\cite{rambleed,yao2020deephammer}. While such a phenomenon was first used by the prior work in~\cite{rambleed} to successfully steal crytpo keys, we note that such technique is ineffective in stealing secrets in bulk and also cannot be applied in the context of DNNs. We, therefore, propose a set of system-level rowhammer optimization schemes that enable \emph{fast and efficient exfiltration} of partial model weights tailored for DNN application platforms. After recovering the partial information at stage-1, the weight search space of a victim model still remains high. For example, even after recovering 90 \% of the bits in a large model like  VGG-11~\cite{simonyan2014very} (i.e., 1056 Million bits for an 8-bit model), the attacker still needs to train the recovered model with limited data to restore the remaining 10 \% bits (i.e., 105.6 million bits). Therefore, to address these additional challenges, \textbf{in the second stage}, 
we propose a novel substitute model training algorithm with \emph{\strain} weight penalty. The purpose of such a loss penalty term is to utilize the recovered partial weight bits for effectively guiding the substitute model training. Subsequently, \dsteal produces a substitute model which is expected to achieve similar accuracy as the victim model with high fidelity. Moreover, the trained substitute model could help to mount strong adversarial input attacks on the victim model as well. We summarize the major contributions of our work as follows:

\begin{enumerate}
    \item For the first time, we investigate a \emph{new model extraction attack} with the assistance of state-of-the-art side channels in memories that can exfiltrate detailed but partial information of DNN model weight parameters.

    \item We develop \hleak, a system attack technique that utilizes fault-based information leakage through rowhammer to efficiently steal partial model weight parameters at scale. To make rowhammer-based side channel applicable and efficient for attacking DNNs, we propose several novel system-level techniques including: \ding{182} using new memory layout for hammering, \ding{183} implementing efficient DNN weight pages massaging through \emph{inference activity anchoring} and \emph{batched page releasing}, and \ding{184} reverse-engineering backend model memory management (in PyTorch) to enable mapping between physical location of leaked bit to logic location in the model.

    \item Leveraging the leaked bit profile from \hleak, we propose a novel training algorithm for the substitute model with \emph{\strain} weight penalty. It first conducts a data filtering process of the leaked bits to construct a profile consisting of projected searching space for each weight, which helps the optimization of substitute model training with reduced optimization space compared with full scale (i.e., no leaked weight info). Then, a \strain penalty term is added to the cross-entropy loss during training, penalizing each weight to converge near the mean of the projected range, for achieving a substitute prototype of the victim DNN model with comparable accuracy and high fidelity.
    
    \item We build an end-to-end \dsteal framework and demonstrate its efficacy on four popular DNN architectures (e.g., ResNet-18/34, Wide-ResNet-28, VGG-11). We also evaluate our attack on three standard image classification datasets (e.g., CIFAR-10, CIFAR-100, GTSRB). For example, the extracted substitute model has successfully achieved more than \emph{90 \%} test accuracy on deep residual networks for the CIFAR-10 dataset. Moreover, \dsteal attack could achieve similar as white-box adversarial input attack performance (i.e., degrading the model accuracy to 1-2 \% under attack).
    
    \item Finally, for the first time, we demonstrate the effectiveness and efficiency of recovering Most Significant Bit (MSB) only from a victim model. With our \dsteal, it can generate effective adversarial examples with similar attack efficacy as a white-box attack (i.e., 0.16 \%).
\end{enumerate}

\section{Background}

\subsection{Model Extraction}
Model extraction is an emerging class of attacks in Deep Learning applications. It jeopardizes the privacy of the deployed victim model by leaking confidential information (i.e., model architecture, weights, biases, etc.). An ideal model extraction attack would expect to extract the exact copy of the victim model. For a task, the input and output pair data ($X,Y$) $\in$ $\mathbb{R}$ can be drawn from the true distribution $D_A$ to train a DNN model $M_\theta$ with parameters of $\theta$. In this work, we designate this model $M_\theta$ as the \emph{victim model}. To extract the exact model, the attacker will attempt to recover a theft model $\hat{M}_\theta$, such that $M_\theta=\hat{M}_\theta$. However, such an identical model (i.e., same architecture and parameter) stealing is practically challenging if not impossible~\cite{jagielski2020high}.

\vspace{1mm}
\noindent \textbf{Algorithm-based Model Extraction.}
To overcome the challenge of stealing an identical DNN model, prior works~\cite{jagielski2020high,chandrasekaran2020exploring,addepalli2020degan} have defined some possible realistic approaches to extract DNN model information. In \Cref{tab:prior_works}, we summarize the prior DNN model extraction works into three major categories. First, in \emph{direct recovery} method, the attacker attempts to reconstruct the victim DNN model using DNN output scores and gradient information. These works~\cite{milli2019model,jagielski2020high,rolnick2020reverse} have solved layer-wise mathematical formulation and internal functional representation to recover weights. In this setting, the goal of the attacker is to create a functionally equivalent model which is given an input $x \in X$, the recovered model $\hat{M}_\theta$ should follow: $M_\theta(x) = \hat{M}_\theta(x)$. This objective is a weaker version of the exact model extraction method. But it remains a difficult route to succeed in model extraction, as prior works~\cite{milli2019model,jagielski2020high,rolnick2020reverse} have failed to show a successful attack for over 2-layer neural network.

In the second approach (i.e., \emph{learning}), papernot et. al~\cite{papernot2017practical} first proposed substitute model neural network training using input and output pairs of a victim DNN model to mount transferable adversarial input attack. In contrast, recent works~\cite{tramer2016stealing,chandrasekaran2020exploring,orekondy2019knockoff,barbalau2020black,nayak2019zero,jagielski2020high,correia2018copycat,pal2019framework} aim to achieve high accuracy or high fidelity on a task using active learning methods. If the attacker prioritizes task accuracy, then the goal is to construct $\hat{M}_\theta$ such that the probability of [$\argmax \hat{M} (x) == y$] (i.e., true label) is being maximized. As for fidelity extraction, given a similarity function S(.), the goal is to construct a model $\hat{M}_\theta$ such that the similarity index $S(\hat{M}_\theta(x),M_\theta(x))$ between the output of the victim and substitute model is maximized. One of the major drawbacks of the learning-based model extraction approach is the high requirement of input query and access to the victim model's output score/predictions.

\begin{table}[ht]
\centering
\caption{Summary of the existing model extraction methods.}
\label{tab:prior_works}
\scalebox{0.8}{
\begin{tabular}{@{}ccc@{}}
\toprule
Type & Attack & Goal \\ \midrule
Direct/Mathematical Recovery &\cite{milli2019model,jagielski2020high,rolnick2020reverse} & Functionally Equivalent \\
Active Learning/Learning &\cite{tramer2016stealing,chandrasekaran2020exploring,orekondy2019knockoff,barbalau2020black,nayak2019zero,jagielski2020high,correia2018copycat,pal2019framework,papernot2017practical} & Task Accuracy/Fidelity \\
Side channel \& Learning &\cite{batina2018csi,hu2020deepsniffer,zhang2021stealing,yu2020deepem,wei2020leaky,duddu2018stealing,244042,xiang2020open}
&  Functionally Equivalent/Fidelity  \\ \bottomrule
\end{tabular}}
\end{table}


\vspace{1mm}
\noindent \textbf{Side Channel Attacks on DNNs.}
There has been a large body of studies on hardware/microarchitecture side channel exploitation where attackers can leak confidential system information through power, EM and timing information on various platforms~\cite{callan2015fase,naghibijouybari2018rendered,zhang2021red,2019cotsknight,fang2019prodact}. Recent works have demonstrated that such attack vectors can also be applied to exfiltrate sensitive DNN information~\cite{batina2018csi,hu2020deepsniffer,zhang2021stealing,yu2020deepem,wei2020leaky,duddu2018stealing,yan2020cache}. 
Among the existing techniques, side channel attack is a more practical strategy to steal sensitive information about a deeper (i.e., many layers) victim model. Usually, the goal of side channel attack is to produce a functionally equivalent model or achieve high fidelity on a dataset. To achieve this, they often supplement side channel attacks with a learning scheme to train a substitute model using the leaked parameter information. This substitute model can later generate adversarial input samples with high transferable properties to attack the victim model more efficiently~\cite{hu2020deepsniffer}. Note that these side channel attacks can only recover model architecture information accurately, but to date, no attacks can steal information about model weights.

In summary, among the three popular directions of model extraction attack, our proposed attack surface falls into the category of side channel attack, leveraging the memory-based fault injection attack rowhammer \cite{kim2014flipping}. Our goal is to use our proposed DeepSteal framework to reproduce a copy of the victim model (i.e., substitute model) with high accuracy and high fidelity on a task. Finally, such a substitute model can later generate adversarial samples with high transferable properties to the victim model.



\subsection{Rowhammer Attacks}
\label{subsec:rowleak}

Rowhammer is a software-based fault-injection attack that exploits DRAM disturbance errors via user-space applications~\cite{kim2014flipping}. Particularly, it has been shown that accesses (e.g., activations) to certain DRAM row can introduce electrical disturbance to the DRAM cells in its neighboring rows, which accelerates the leakage of their charges in the capacitors~\cite{kim2014flipping,mutlu2019rowhammer}. An attacker can intentionally activate one DRAM row (whose data belongs to the attacker) frequently enough (i.e., hammering) that will eventually cause certain cells in neighboring rows to lose sufficient charge, leading to bit flips in memory. Such attacks have been successfully demonstrated on commercial-off-the-shelf DRAM modules even with the presence of ECC features~\cite{another_flip,eccploit}. There are mainly three hammering techniques proposed in the literature: a) double-sided hammering~\cite{gruss2016rowhammer,mutlu2019rowhammer,kim2014flipping,seaborn2015exploiting,eccploit}: where two adjacent aggressor rows to the target row creates a sandwich pattern with target row as inner layer and the aggressors rows repeatedly activated; 
b) single-sided hammering~\cite{seaborn2015exploiting}: where one adjacent row to the target row and another random row are activated repeatedly; and c) one-sided hammering~\cite{another_flip}: where one adjacent row is repeatedly activated maintaining a fixed frequency to exploit \textit{close-page} DRAM policy (i.e., system policy that automatically deactivates the activated row if not accessed within a determined time frame). 
Among all those techniques, double-sided hammering is the most effective in inducing DRAM faults since it introduces the strongest disturbance. 




\begin{figure}[t]
    \centering
    \includegraphics[width=0.44\textwidth]{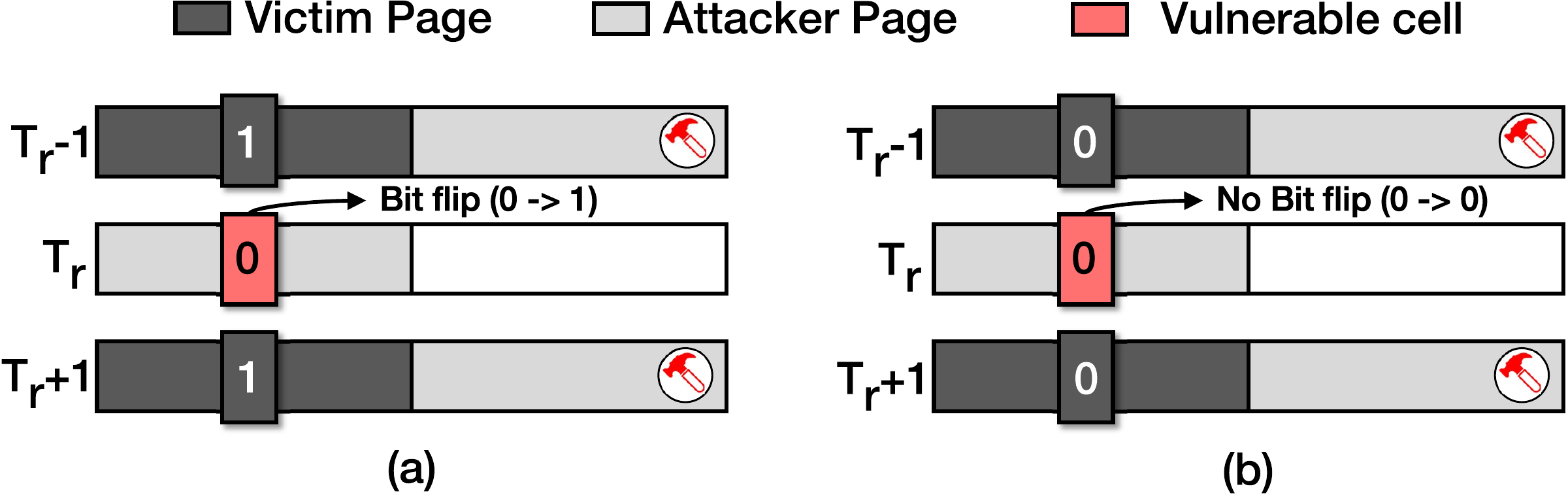} 
    \caption{Data dependency for inducing a rowhammer fault. Here, based on the presence of bit flip in the \textit{attacker-controlled} vulnerable bit in target row ($T_r$), data from adjacent row from victim program can be inferred. }
    \label{fig:rambleed}
\end{figure}


There are a large body of works demonstrating many variants of Rowhammer attacks. Most of them focus on tampering with the integrity of victims, including privilege escalations~\cite{seaborn2015exploiting}, system denial of service~\cite{jang2017sgx} and more recently faulting DNN model parameters~\cite{yao2020deephammer,caiseeds}. Several works also use rowhammer as the fault injection attack vector to recover crypto-keys through differential fault analysis~\cite{flip-fenshui}. Until recently, RAMBleed~\cite{rambleed} reveals that the rowhammer fault characteristic can be leveraged to carry out \emph{information leakage} attacks that directly infer information about certain victim secret. This attack leverages the fact that a column-wise data dependence is required to successfully flip a bit for a \emph{known vulnerable memory cell}. Notably, under the most effective double-sided hammering, a bit flip for a vulnerable cell can succeed with high confidence if its upper and lower bits in the same column (e.g., in the aggressor rows) store the opposite bits (\texttt{1--0--1} or \texttt{0--1--0}), or fail if such pattern is not in place. \Cref{fig:rambleed} illustrates such a data dependency for cell with bit flip vulnerability (in the $0\rightarrow1$ direction). As we can see, for the victim page in the middle with a vulnerable bit set to `0', a bit flip would only occur if the direct top and bottom bits are set to `1s', thus achieving the column-wise striped pattern. In RAMBleed, the attacker puts his own page in the middle row that has the vulnerable cell, and manages to trigger the placement of two copies of victim's secret in the corresponding aggressor rows. By observing whether a bit flip occur in his own page after hammering, the attacker can infer the secretive bit (i.e., RSA keys) from the victim stored in the aggressor rows. We note that exploitation of rowhammer in the domain of information opens new direction in terms of information security beyond the scope of integrity tampering. 

\




\color{black}



\section{Threat Model}
The attacker targets on exfiltrating internal information (i.e., model weights) from deep learning systems by exploiting the underlying hardware fault vulnerabilities in modern computing systems. The attacker can control a user-space un-privileged process that runs on the same machine as the victim DNN service. We assume that the deep learning system is deployed on a resource-sharing environment to offer ML inference service. Such application paradigm is becoming popular due to the prevalence of machine-learning-as-a-service (MLaaS) platforms~\cite{ribeiro2015mlaas}. Our proposed framework manifest as a \emph{semi-black box} attack where the adversary does not have any prior knowledge of the model weights. However, the attacker is aware of some key model architecture information, including model topology and layer sizes. We note that such an assumption is legitimate, as prior works demonstrate many practical ways to recover model architecture information through various side channel exploitation (e.g., through caches~\cite{nas}, memory bus~\cite{hu2020deepsniffer} and EM~\cite{yu2020deepem}). 


In this work, we leverage the rowhammer fault attack vector that commonly exists in today's DRAM-based memory systems (i.e., DDR3, DDR4) as the side channel~\cite{rambleed}. Specifically, the attacker takes advantage of the fact that bit flip in \emph{vulnerable DRAM cells} only occur when the column-wise bit striping pattern exists. By leveraging such data dependency, the attacker can infer bits in the aggressor rows by observing if a bit flip occur in \emph{his/her own address space} (i.e., in the middle row). In other words, the attacker do not directly tamper with the victim's memory (as shown in most traditional rowhammer attacks), therefore it remains stealthy as it does not incur system-level alarming event (e.g., crashes). The attacker may share certain read-only memory together with the victim DNN (e.g., ML platform binaries) either through library sharing or advanced memory deduplication feature supported in modern OS~\cite{caco}.
Our attack mainly utilizes double-side rowhammer since fault occurrence in such setup exhibits stable data dependency. 
We assume that proper confinement mechanism is put in place to disallow direct access of data across processes. We further assume that the operating system and the hypervisor are benign and appropriate kernel-space protection mechanism are deployed to avoid direct tampering of kernel structures~\cite{konoth2018zebram}.


For substitute model training, as depicted in \Cref{tab:threat_white}, we assume the attacker has no knowledge of gradients, and is denied access to DNN output scores/predictions, which is more strict than prior works~\cite{chen2017zoo,papernot2017practical}. Meanwhile, following most recent related works ~\cite{cubuk2020randaugment,cubuk2019autoaugment}, the attacker has access to a publicly available portion (e.g., $\leq$ 10 \%) of the labeled training dataset.

\begin{table}[ht]
\centering

\caption{\emph{A list of information accessible to the attacker for substitute model training at stage-2 (\Cref{fig:overview}).}}
\label{tab:threat_white}
\resizebox{0.45\textwidth}{!}{%
\begin{tabular}{@{}lcc@{}}
\toprule
\multicolumn{1}{c}{Attacker Information} & Accessible \\ \midrule
1. DNN Architecture & \checkmark   \\
2. \hleak recovered weight bits &  \checkmark \\
3. Gradient Computation &  \xmark \\
4. Train/Test Data & \xmark    \\
4. Victim model Output & \xmark    \\
5. A portion of publicly available data ($\leq$ 10 \%) & \checkmark   \\ \bottomrule
\end{tabular}%
}

\end{table}

\begin{figure}[ht]
    \centering
    \includegraphics[width=0.5\textwidth]{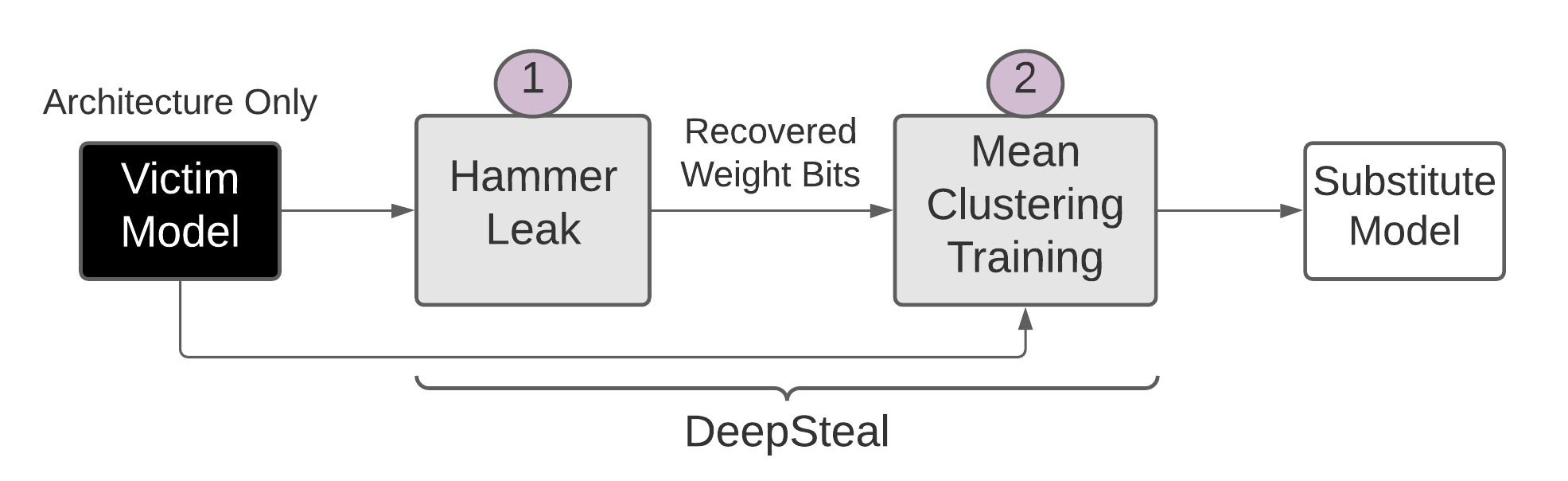} 
    \caption{\textbf{Overview} of our \emph{\dsteal} attack framework. \emph{stage-1:} \emph{\hleak} is a side channel attack to leak sensitive weight bit information through exploiting hardware fault vulnerabilities using rowhammer. \emph{stage-2:} To use the recovered bit information efficiently, we train a substitute model using \emph{\strain} weight penalty and recover a substitute proto-type of the victim model}
    \label{fig:overview}
\end{figure}

\begin{figure}[t]
    \centering
    \includegraphics[width=0.44\textwidth]{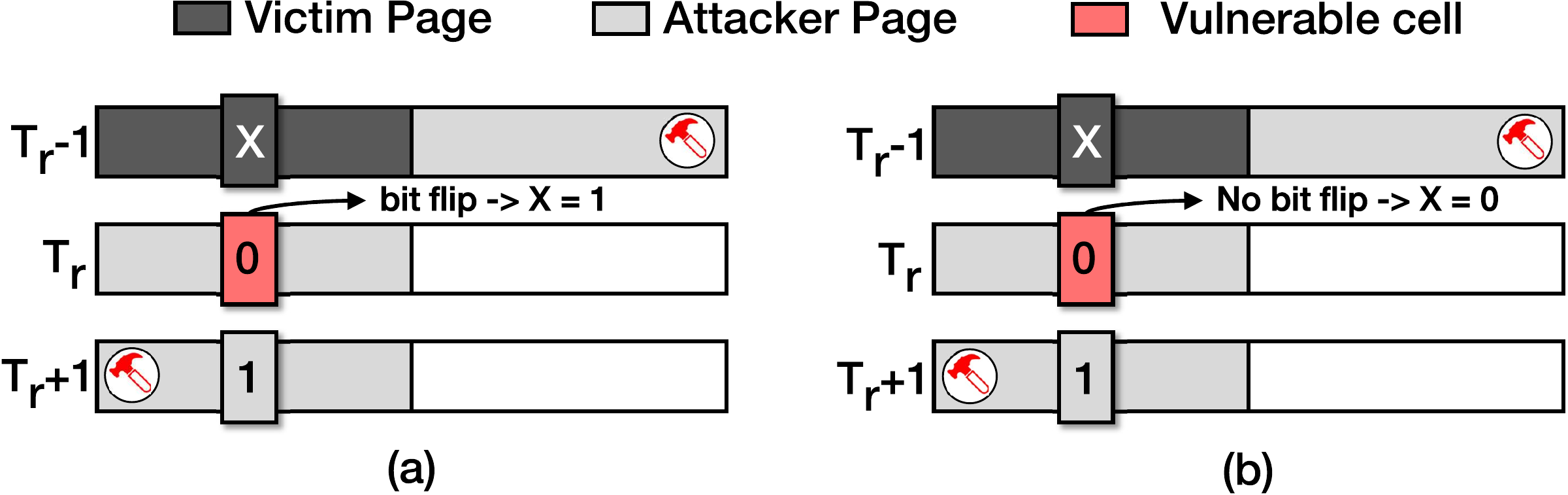} 
    \caption{\hleak attack leaking victim secret using single victim page. (a) Bit flip observed when victim's bit is \texttt{1}, (b) Bit flip not observed when victim's bit is \texttt{0}. }
    \label{fig:hammerleak}
\end{figure}

\begin{figure}[t]
    \centering
    \includegraphics[width=0.485\textwidth]{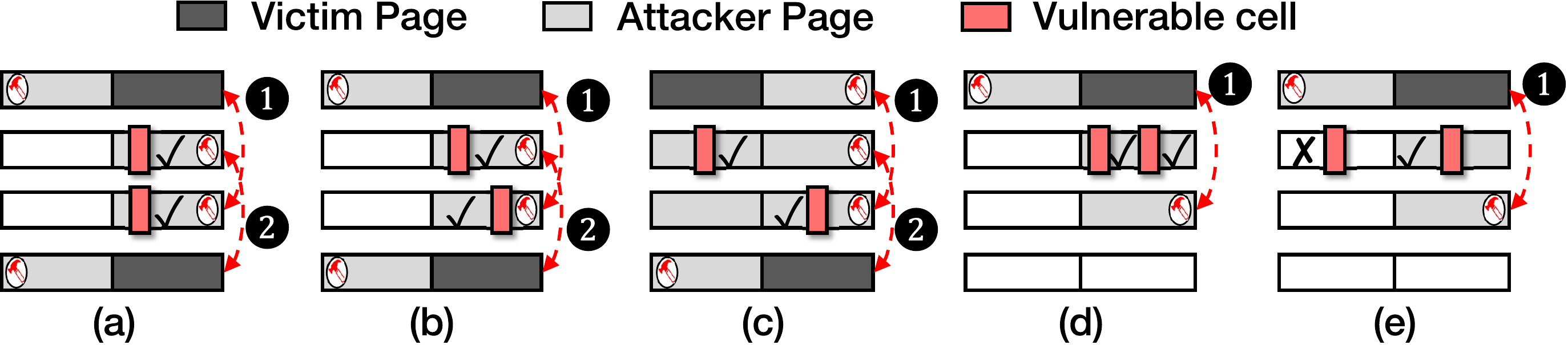} 
    \caption{Efficient utilization of target row holding vulnerable bits. \hleak can utilize all $V_c$ in (a) Adjacent $T_r$ with same $V_c$ location, (b) Adjacent $T_r$ with different $V_c$ location (mirror page), (c) Adjacent $T_r$ with different $V_c$ location (alternate page), and (d) Same $T_r$ with different $V_c$ location (same page). Only in the case of (e) Same $T_r$ with different $V_c$ location (different page), one page in $T_r$ can be used.}
    \label{fig:hammerleak_char}
\end{figure}

\section{Overview of \dsteal}

In this work, we propose a novel adversarial model stealing mechanism for Deep Neural Networks through rowhammer. An overview of our proposed attack framework, \emph{\dsteal}, is presented in \Cref{fig:overview}. It has two key components: i) an efficient rowhammer-based weight-stealing module \emph{\hleak} \& ii) a substitute model training mechanism with novel \emph{\strain} loss penalty. At stage-1 in \Cref{fig:overview}, we mount \hleak attack on inference infrastructure (i.e., a remote machine running on-demand inference using the DNN model) to recover weight bits corresponding to each layer of the model (more explanation about \hleak is discussed in \Cref{sec:hammerleak} and case study in \Cref{subsec:case_study}. We continue the \hleak for many rounds until the desired portion of weight bits are recovered.

Once a portion of the neural network weight bits are leaked using \hleak, at stage-2, our goal is to use these leaked weight bit information and generate a substitute prototype of the victim model. To achieve this, we propose a novel neural network training algorithm with \strain loss penalty on the weights to constrain the trained substitute model weight parameters to be as close as possible to the recovered partial weight info, as well as minimizing the accuracy loss. 
The learned substitute model will pose the following properties: i) it has a similar test accuracy as the victim's model; ii) high fidelity; \& iii) the attacker can use this substitute model to generate strong adversarial input samples to attack the victim model with higher efficacy.
Next, we will describe the details of stage-1 and stage-2.

\section{\hleak: Efficient rowhammer based Data Stealing}
\label{sec:hammerleak}

In this section, we present the \textit{\hleak} framework, which is built on advanced rowhammer-based side channels to leak secretive data (model weigt bits) from the memory of victim DNN applications in bulk.

\subsection{Rowhammer Information Leakage for DNNs} 
\label{subsec:hammerleak}

Though the rowhammer based attack in RAMBleed (\cite{rambleed}) demonstrates to steal information from the aggressor rows adjacent to the vulnerable cell, it requires the same page content in both aggressor rows. While it is possible that certain applications may launch multiple threads with each allocating memories to store a copy of secrets (e.g., as in the case of OpenSSH shown in~\cite{rambleed}), we find that such duplication does not exist in modern machine learning frameworks (e.g., pytorch~\cite{pytroch}). As such, the RAMBleed-style leakage manifest does not work for DNNs. To address this issue, we augment rowhammer by exploiting the same observation of data dependecy, but with added capability of leaking victim bits with only one copy of victim page, which enables the rowhammer based information leakage applicable to any victim application. \Cref{fig:hammerleak} illustrates the major design of augmented rowhammer leakage. Unlike typical rowhammer based information leakage attack, the improved rowhammer leakage does not require two duplicated copy of the victim page, instead it substitutes one victim page with attackers page while still being able to leak secret bits from the victim page. To carry out the attack, the attacker first places the victim page to one of the adjacent aggressor rows (e.g., top aggressor row, $T_{r-1}$) and then places attacker's own page in the other aggressor row (e.g., bottom aggressor row, $T_{r+1}$) and the target row ($T_r$), as shown in \Cref{fig:hammerleak}. The content of the attacker rows are controlled such that it creates a bit layout of \texttt{X--0--1} where \texttt{X} is the secretive victim bit, the middle bit is the $V_c$ and the last one is another attacker controllable bit. Now, if the two aggressor rows are hammered, a bit flip will be observed in $V_c$ if \texttt{X} is \texttt{1} as shown in \Cref{fig:hammerleak}a (thus creating a \texttt{1--0--1} pattern), otherwise there will be no bit flips in $V_c$ (\Cref{fig:hammerleak}b). This same attack can also be carried out considering \texttt{X--1--0} bit layout, using the same technique. 

\begin{figure}[t]
    \centering
    \includegraphics[width=0.36\textwidth]{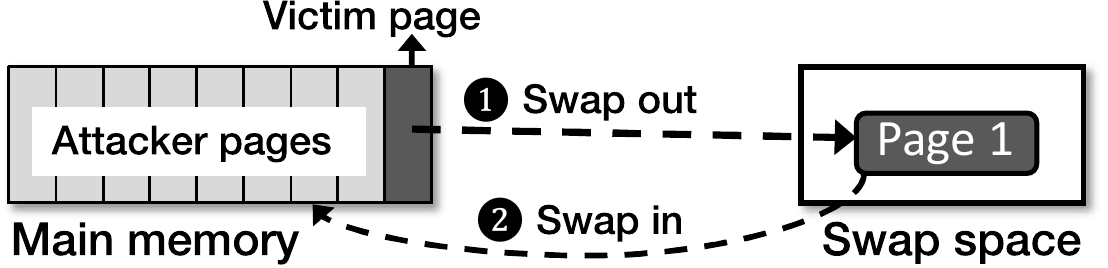} 
    \caption{Memory exhaustion technique to occupy large memory area by attacker, forcing victim data to \textit{swap out} (\ding{182}). When victim accesses the data, its \textit{swap in} (\ding{183}) to a different memory location}
    \label{fig:enhaust}
\end{figure}

\begin{figure*}[t]
    \centering
    \includegraphics[width=0.80\textwidth]{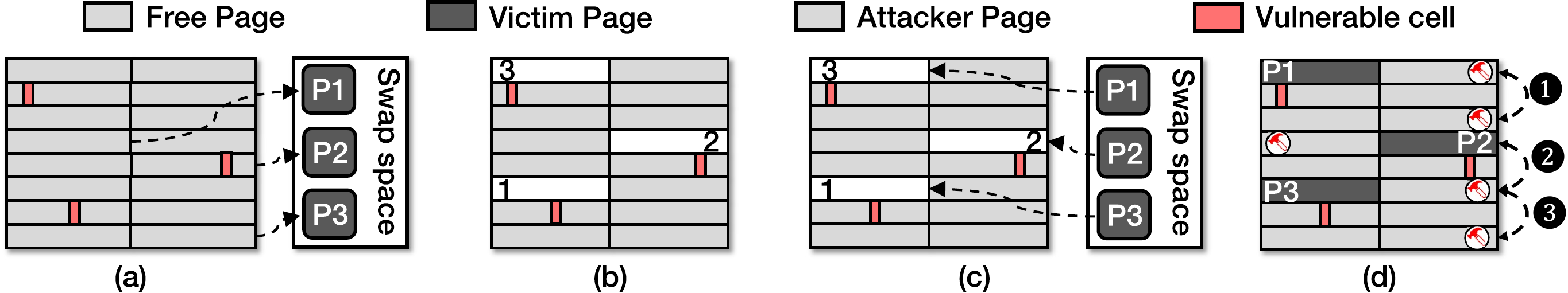} 
    \caption{Overview of the operations in the \hleak fraemwork.}
    \label{fig:hammerleak-attack}
    \vspace{-3mm}
\end{figure*}
\vspace{1mm}
\noindent\textbf{Efficient Utilization of Vulnerable Rows.} As the vulnerable DRAM cells are typically randomly distributed across a DRAM DIMM~\cite{hammertime}, it is possible that multiple vulnerable cells are present in the same row or in adjacent rows. In that case, the improved rowhammer leakage can create smart memory layout to increase the utilization of vulnerable bits in leaking secrets. \Cref{fig:hammerleak_char} enumerates all possible cases of multiple $V_c$ in same or adjacent rows. In case of multiple $V_c$ in adjacent rows (i.e., \Cref{fig:hammerleak_char}a, \Cref{fig:hammerleak_char}b and \Cref{fig:hammerleak_char}c), we can setup the memory layout in such a way that it first leaks secret from first $T_r$ (\ding{182}) by using the second $T_r$ as an aggressor, then leaks from the second $T_r$ by using the first $T_r$ as an aggressor (\ding{183}). In case of multiple $V_c$ in the same page (i.e., same $T_r$ as in \Cref{fig:hammerleak_char}d), we can leak multiple secret bits at the same time by setting proper bit layout in the $T_r$ and attacker-controller aggressor. Only in case of different $V_c$ is different page of the same $T_r$, we can use only one page at a time since the attacker needs to control a page in the aggressor row containing victim page in order to carry-out the hammering (\Cref{fig:hammerleak_char}e). In general, in a DRAM row with $n$ individual pages, we can leak from upto $n-1$ pages while keeping one page as row activation page used for hammering. 
Note that this is considering a system with a \textit{single-channel} memory which contains two 4KB pages per physical row. In systems with \textit{dual-channels}, even more bits could be leaked per physical page. 


\subsection{Victim Secret Leakage using \texttt{\hleak}}

At a high level, \hleak is a multi-round attack framework. In each round the attacker first tweaks the placement of victim pages in appropriate location (i.e., adjacent to $T_r$) using memory exhaustion and deterministic victim relocation via Bitflip-aware page release and then recovering the secret bits using improved rowhammer leakage. We now discuss each of these steps in details.

\vspace{1mm}

As rowhammer bitflips are highly dependent on specific DRAM locations, the victim pages containing secret must be placed appropriately to create desired memory layout with vulnerable cells (i.e., $V_c$). This is challenging because: i) typically $V_c$ locations are sparse and only a few bits can be leaked in each round of hammering. Hence, to leak \textit{new} bits, the victim page must relocate in different physical region which is not in direct control by the the attacker. Note that prior memory massaging techniques such as page deduplication~\cite{flip-fenshui} or page cache evictions~\cite{another_flip} do not work as we find out DNN platforms such as pytorch allocate anonymous pages for DNN model weights; and 
ii) the physical address of victim page is not accessible by an attacker running as user-space process. 
We propose a four-step \hleak framework tackling each of these challenges to augment a practical rowhammer based side channel.


\begin{enumerate}[label=\textbf{Step \arabic{enumi}: }, leftmargin=0cm,itemindent=.5cm,labelwidth=\itemindent,labelsep=0cm,align=left,itemsep=2pt]

\item \textbf{Anonymous Page Swapping.} During this step, the attacker's goal is to evict victim application pages from main memory to swap space so that they later get relocated by the operating system the next time they are accessed by the victim. This procedure is illustrated in \Cref{fig:enhaust} where the victim page \textit{Page 1} is \textit{swap out} of main memory to the swap space because of attacker occupying the physical memory (\ding{182}). Once the victim application accesses this page again, the OS \textit{swaps in} \textit{Page 1} to main memory in a new location (\ding{183}). This is the first step towards the attacker's goal of relocating the victim page in a new location so that he can leak new bits. To achieve this, the attack allocates large memory using \texttt{mmap} with \texttt{MAP\_POPULATE} flag. This triggers the OS to move other data (including victim application pages) out of main memory to swap space. After the end of this phase (\Cref{fig:hammerleak-attack}a), the attacker occupies most of the physical memory space with victim pages stored in swap space.

\item \textbf{Bitflip-aware Page Release.} After the memory exhaustion where the attacker process occupies the majority of the system memory, the attacker systematically releases pages for victim pages to relocate into. Given the knowledge of virtual to physical page frame mapping of attacker process (using the same memory massaging technique as memory template), the attacker first creates a list of potential pages for the victim to occupy as \textit{aggressor} during rowhammer attack (i.e., pages that are adjacent to a $T_r$ holding one or more $V_c$). During each round of \hleak, the attacker randomly chooses a predetermined number of pages from the list. Finally, the attacker releases the selected pages (i.e., by calling \texttt{munmap}) to release the pages (\Cref{fig:hammerleak-attack}b). Note that the ordering of page release (i.e., \texttt{1,2} and \texttt{3} in that order as illustrated in \Cref{fig:hammerleak-attack}c) is critical in this step as we discuss next.

\item \textbf{Deterministic Victim Relocation.} The main goal of this step is to place victim pages in a predetermined location which is optimized to create appropriate memory layout for rowhammer leakage. In addition, this also ensures that the victim page location is known to the attacker so that the attacker can correlate leaked secret with exact data from the victim domain. For this purpose, we utilize the per-core page-frame cache structure called \texttt{per-cpu pageset}~\cite{gorman2004understanding}. This structure is maintained by the Linux kernel and is used to hold recently released pages by that CPU core. This is a stack-like \textit{last-in-first-out} (LIFO) structure that holds the recently freed pages by processes running on that core. When the OS needs to allocate a page for a process, it first checks the pageset corresponding to the physical core to obtain a free memory location. Note that this structure is shared between hyperthreads running on the same CPU core, hence an attacker can exploit the LIFO policy of pageset to deterministically place victim pages into the desired memory location by running on the same CPU core (either in round-robin scheduler or in simultaneous multithreading). In particular, based on the order of release of pre-selected pages by the attacker during \textit{Step 2}, the location of victim page placements can be determined with extremely high accuracy~\cite{yao2020deephammer} as illustrated in \Cref{fig:hammerleak-attack}c. Here \texttt{P1} (page 1), \texttt{P2} and \texttt{P3} are allocated for the victim process in that order and these pages are placed in memory location \texttt{3,2} and \texttt{1} respectively following the reverse order of page release during Step 2.

\item \textbf{Recovering Secrets Using Rowhammer.} After completing Step 3 in which the victim pages are placed in appropriate location, the attacker mounts our rowhammer-based side channel (as discussed in \Cref{subsec:hammerleak}) to steal victim data.
Based on the flip direction of a specific $V_c$ (i.e., flip in either \texttt{0$\rightarrow$1} or \texttt{1$\rightarrow$0} direction), we preset the $V_c$ and adjacent attacker controlled aggressor row bit respectively to \texttt{0--1} (for \texttt{0$\rightarrow$1}) or \texttt{1--0} (for \texttt{1$\rightarrow$0}). After hammering, the attacker reads the value of $V_c$ and examines for bit flips. An observed bit flip indicates that the adjacent bit in victim controlled aggressor row is the same as the preset bit in the attacker's controlled aggressor row. This way, the attacker can recover secret bits from all of the selected $T_r$ iteratively to maximize the data leakage in one round of \hleak.


\end{enumerate}




These four steps correspond to one round of \hleak attack on the victim application. \hleak is a chained operation where the attacker resumes the attack again to leak different bits and this chained operation is continued until the attack goal (e.g., a specific page recovery) is reached. Note that the attacker must choose different sets or different orders of targeted page (Bitflip-aware) for release in Step 2 in order to put victim pages into locations where it is possible to leak more bits. Since the attacker has fine-grained controlled over the placement of victim pages, he can optionally optimize the pages to release in \textit{Step 2} depending on which data is yet to recover, instead of relying of randomly releasing pages. For example, if the attacker needs to recover bit offset \texttt{x} from a victim page, he can chose to release a page which is adjacent to a $T_r$ with $V_c$ in bit offset \texttt{x}. This greatly increases the efficiency of \hleak by reducing the number of rounds to execute to recover certain number of bits.

\subsection{Batched Victim Page Massaging}

While the one-time \textit{bitflip-aware page release} for one round of \hleak is sufficient for victim applications with small memory footprint (i.e., the size of working set is less than 512 pages), this is not true for larger applications since per-cpu pageset has a fixed size.
For applications that have large working set (i.e., larger than 2MB), deterministic victim page relocation cannot be guaranteed if all required pages by victim in one round is released at once during Step 2. This is because if the required number of pages exceeds the size of pageset, the pageset is overflown and it starts to release some of the free pages to global memory pool. At that point, we can no longer maintain the LIFO ordering of pageset, which could lead to the failure of deterministic victim page relocation. 

We propose \textit{batched victim page massaging} to address this issue, which periodically releases a small number of pages at specific point(s) of victim execution so that the pageset does not get overflown. Additionally, this allows the attacker to have better control over filtering non-secretive victim pages by placing them at unleakable locations. To determine \textit{when} to release leakable pages, the attacker needs to monitor victim's activities that lead to secretive page accesses. To simplify this process, we assume that both the attacker and the victim application use the same library (which results in shared read-only physical pages~\cite{caco}). 
For applications using open-source library (which is the case for majority of the cryptographic applications and machine learning applications), the attacker can analyze and determine the secret accessing execution flow through source/binary code examination (e.g., a specific function which performs computation using secretive data), otherwise the attacker needs to perform binary analysis. Then the attacker instantiates FLUSH+RELOAD at cerntern \emph{anchor point} in that function~\cite{yarom2014flush+}\footnote{The attacker repeatedly flushes the cacheline containing the function code and times the subsequent access to the same function. A short access time means victims has accessed this function, otherwise the victim did not access.}. We build a cache side channel based monitoring tool to setup function anchor point for tracking purpose. If library sharing is not possible, we can alternatively use PRIME+PROBE based anchors.
Additionally, through experimentation or source code analysis, the attacker pre-determines how many non-secretive pages the victim allocates after accessing the anchored function and what is the victim access pattern (e.g., if there are non-secretive page access between a chunk for secret page access) and uses that information in releasing pages.

Putting everything together, \textit{prior to the attack}, the attacker determines an anchor point in victim's execution path which symbolizes the victim accessing certain secret. The attacker also determines secret page access pattern (i.e., number of non-secretive pages before accessing the secret, $P_{b}$, secret page accesses, $P_{s}$, and intermediate non-secretive pages access between secrets, $P_{i}$). 
\textit{During the attack phase}, the adversary monitors accesses to the anchor points. Once the victim access is detected, the attacker releases $P_b+P_s+P_i=P$ physical pages considering the reverses order of victim accesses (i.e., LIFO access). Out of these pages, only $P_s$ pages need to be the victim's pages for information leakage. If $P$ is larger than the size of per-core pageset, the attacker will choose a different anchor point, which effectively divides the victim secret page accesses to additional smaller chunks. By chaining several page batches for memory massaging, \hleak can eventually steal bits spanning \emph{hundreds to thousands pages} in each individual round, which enables weight leakage for DNN applications with large memory footprint. 




\section{Substitute model training with \strain}


At Stage-2 of \dsteal, we will leverage the bit information leaked by \hleak to learn a substitute prototype of the victim DNN model. The challenge is that the recovered weight bits are not complete, with a mixture of different significant bits. To fully leverage those leaked partial bit-wise data, we propose a novel substitute model training algorithm with a mean clustering loss penalty to reconstruct a neural network model, targeting for high accuracy \& high fidelity. Moreover, this learnt substitute model will help the attacker generate highly effective adversarial input samples to fool the victim model successfully.

\begin{figure}[ht]
    \centering
    \includegraphics[width=0.499\textwidth]{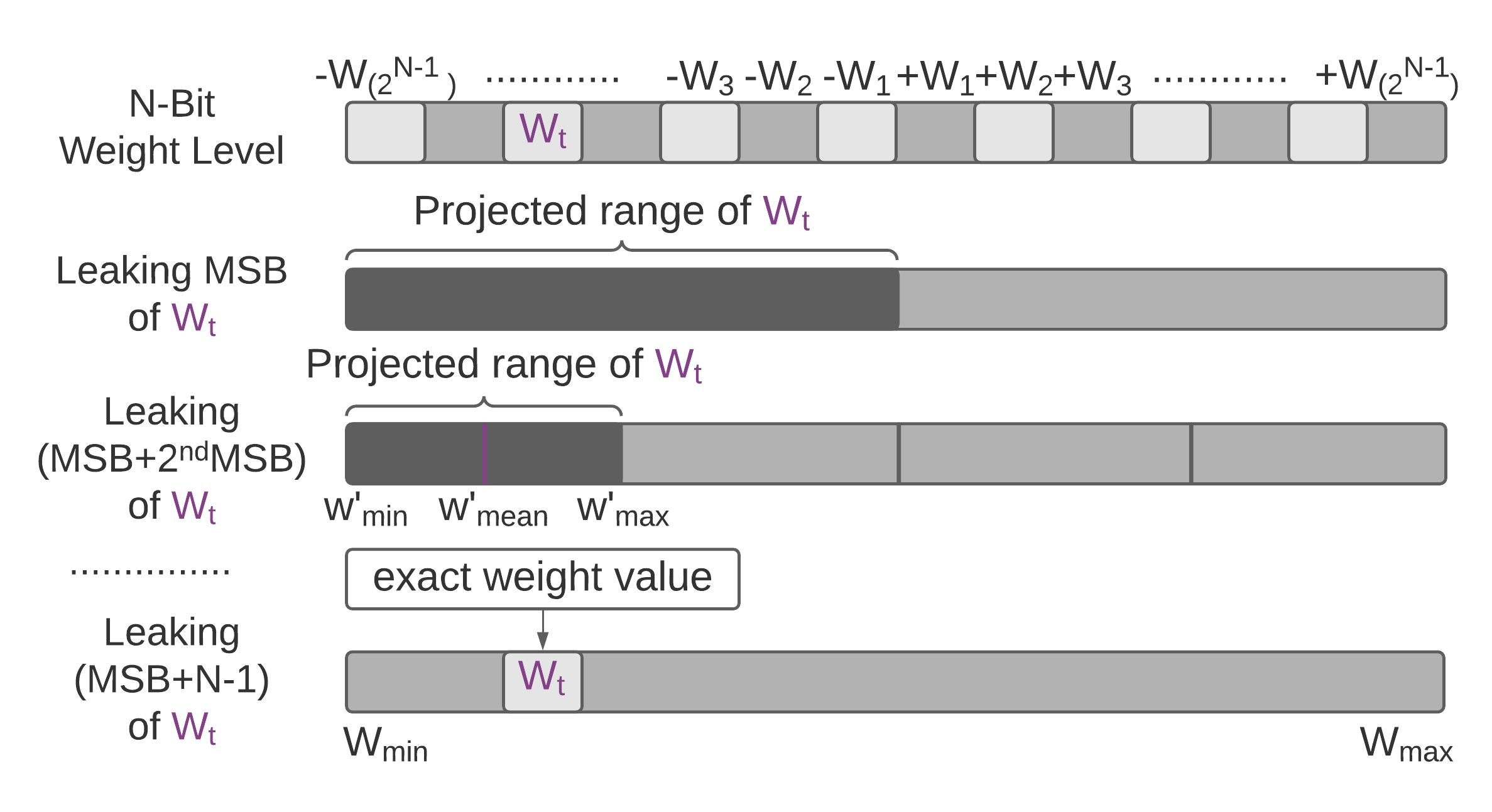} 
    \caption{\emph{First row:} N-Bit quantized weight level; \emph{Second row:} Once the MSB of a weight $W_t$ in the victim model is leaked, we can narrow down the projected range of $W_t$ in the substitute model; \emph{Last row:} Leaking all the bits can track down the exact value of $W_t$ for the substitute model training.}
    \label{fig:algo}
\end{figure}

\subsection{Hammer Leaked Data Filtering}


At stage-1, \hleak recovers a portion of the neural network weight bit information scattered across different significant bits (i.e., from LSB to MSB) for each weight. However, not all of those recovered bits will be used for substitute model training, since it is a mixture of significant bits for each weight. As shown in Fig.\ref{fig:algo}, for each weight, it is preferred to recover MSB first. Thus it forms a smaller and closed searching space (i.e., either positive or negative), rather than full-scale space, for this weight during substitute model training to minimize loss. With the knowledge of MSB, the $2^{nd}$ MSB or more following bits will further reduce the closed searching space. Otherwise, only recovering lower significant bits, without higher significant bits, does not provide much useful information about this weight's potential range. As a consequence, to use the leaked bit information effectively from stage-1 of \hleak, the attacker must filter and reorganize the leaked bits in a sequence from MSB to LSB. Therefore, in this work, before substitute model training, we sort out the leaked weight bits in the following sequence: MSB leaked, MSB + $2^{nd}$ MSB leaked, MSB + ($2^{nd}$ \& $3^{rd}$ MSB) leaked, and so on, to develop a profile for each weight with projected range, as described in Fig.\ref{fig:algo}. Note that, if no MSB is recovered, the projected weight value range will be treated as full scale. 

In \Cref{fig:algo}, we visualize the relationship between i) filtered bits (leaked by \hleak) information of weight from a victim model and ii) the expected range of that corresponding weight during the training of the substitute model. It shows gradually leaking more bit information (i.e., MSB, MSB+$2^{nd}$ MSB,..) of one target weight $W_t$ can help an attacker reduce the searching space of $W_t$ during model training. We define this expected range as \emph{projected range} of each weight in the substitute model.

\subsection{\strain Optimization}
Leveraging the profile of such projected range of each weight, we propose a novel training algorithm for the substitute model using \textit{\strain} weight penalty. It applies an additional loss penalty to the cross-entropy loss during the training process. The \strain penalty term aims to penalize each weight to converge near the mean of the projected range. 

To formally define the problem, let's consider the weight matrix of the victim DNN model at layer $l$ to be $\bm{\hat{W}}^l$. Based on the leaked weight bits at this layer, the attacker can compute the projected range of each weight in the substitute model $\bm{W}^l$. The projected range can be represented as: $\bm{W}^l_{min}$ \& $\bm{W}^l_{max}$ matrix; the minimum and maximum projected value matrix corresponding to each weight in $\bm{W}^l$. Using this closed range, the projected mean matrix $\bm{W}^l_{mean}$ is computed as: ($\bm{W}^l_{max}$ + $\bm{W}^l_{min}$)/2. Next, leveraging this mean matrix, we propose to design a \strain loss penalty as highlighted in \Cref{eqt:piecewise_clustering}. This loss term is added to the inference loss $\mathcal{L}$ and the  optimization process can be formulated as:

\begin{equation}
\label{eqt:piecewise_clustering}
\begin{gathered}
    \min_{\{\tW_l\}_{l=1}^L} \mathbb{E}_{\vx}~\mathcal{L}(f(\vx,\{\tW^l\}_{l=1}^L), \vy) + \\ 
    \underbrace{\lambda \cdot \sum_{l=1}^{L}(||\tW^l - \tW^l_{mean}||) }_\textrm{loss penalty for \strain}
\end{gathered} 
\end{equation}

Here, $\lambda$ is a hyper-parameter that controls the strength of the loss penalty and f(.) denotes the inference function of the DNN model for an input-label pair $(\vx,\vy)$. 
The first term of the loss function in \Cref{eqt:piecewise_clustering} is a typical cross-entropy loss for neural network training using gradient descent. The purpose of the additional \strain loss penalty is to penalize each weight to converge near $\{\bm{W}^l_{mean}\}_{l=1}^L$. 

\begin{algorithm}[t]
\caption{\emph{Substitute Model Training with \strain}} \label{alg:algorithm}
\begin{algorithmic}[1]
\Procedure{Train Substitute Model ($M_\theta$,$\hat{M}_\theta$,$\hat{\theta}_{l=1}^L$)}{} 
\State Takes victim model architecture $\hat{M}_\theta$ as input
\State Takes the leaked parameter list $\hat{\theta}_{l=1}^L$ as input.
\State Randomly Initialize Substitute model $M_\theta$.
\State Perform step-1, data filtering of the leaked bits.
\For{Each layer $l = 1, \ldots, L$}
\State Compute Initial $\bm{W}^l_{min}$ \& $\bm{W}^l_{max}$ using $\hat{\theta}_l$.
\State Estimate $\bm{W}^l_{mean}$ = ($\bm{W}^l_{min}$+ $\bm{W}^l_{max}$)/2.
\EndFor

\State

\State Re-Initialize model $M_\theta$ using following rules:
\State \emph{\textbf{Weight Set-1 (Full 8-bit recovered)}}: Initialize the  \hspace{3em}-\hspace{1.3em}weights at the exact recovered value. This weight set \hspace{3em}-\hspace{1.3em}will not be trained (i.e., set gradients to zero).

\State \emph{\textbf{Weight Set-2 (Partial bit recovered (i.e., MSB+n; \hspace{3em}-\hspace{1.2em} n = 0, $\ldots$, 6))}}:  Initialize  the weights using the \hspace{5em}-\hspace{1.2em}  $\{\bm{W}^l_{mean}\}_{l=1}^L$ matrix \& set $\lambda$ suitably.

\State \emph{\textbf{Weight Set-3 (0 bit recovered)}}: Random Initialization \hspace{3em}-\hspace{1.2em} \& set $\lambda$ in \Cref{eqt:piecewise_clustering} to zero.
\State
\For{each training iteration}
\For{each training batch $(\vx,\vy)$}
\State Compute Loss using \Cref{eqt:piecewise_clustering}.
\State Perform a gradient descent step to update $\theta$.
\State Update $\bm{W}^l_{min}$, $\bm{W}^l_{max}$ \& $\bm{W}^l_{mean}$.
\EndFor
\State Clip $\{\bm{W}^l\}_{l=1}^L$ within the projected range. 
\EndFor
\State \textbf{Return:} Trained Substitute Model $M_\theta$.
\EndProcedure
\end{algorithmic}
\end{algorithm}

\subsection{Overall Training Algorithm.}
We summarize our proposed training algorithm in \Cref{alg:algorithm}. After the filtering step, we divide the weights into three categories: \emph{Weight Set-1:} Full 8-Bit recovered, \emph{Weight Set-2:} Partial bit recovered (i.e., MSB + n; n=  0,...,6) \& \emph{Weight Set-3:} No bit recovered. For set-1, the attacker knows the exact weight value in the victim model. Hence, we will use the exact recovered value for the substitute model by freezing (i.e., set gradient to zero) them during training. The second set of weights is trained using the proposed loss function in \Cref{eqt:piecewise_clustering}. And for set-3, we do not apply the \strain loss penalty (i.e., $\lambda$=0). Both set-2 \& set-3 weights are trained using standard gradient descent optimization. During training, each time before computing the loss function in \Cref{eqt:piecewise_clustering}, we update the projected mean matrix $\{\bm{W}^l_{mean}\}_{l=1}^L$ using the weights of current iteration. If any weight value exceeds the projected range, it will be clipped. Finally, for the last few iterations (e.g., 40), the model will be fine-tuned (i.e., $\lambda = 0$, no clipping \& low learning rate) to generate the final substitute model.

\section{Experimental Setup}

\subsection{Attack Evaluation Metrics}
To evaluate the efficacy of our \dsteal attack, we adopt three different evaluation matrices, i.e., accuracy of the substitute model, fidelity of the substitute model, and accuracy of the victim model under adversarial input attack.
 
\paragraph{Accuracy (\%)} It is the measurement of the percentage of test samples being correctly classified by the substitute model for a given test dataset. Note that, this is the same test data used for victim model. For an ideal successful model extraction attack, we expect the accuracy of the victim and substitute model to be almost identical.
\paragraph{Fidelity (\%)} We measure the \emph{fidelity} as the percentage of test samples with identical output prediction label between the victim model and substitute model. This follows the definition of~\cite{jagielski2020high}, where two models with high fidelity should agree on their label prediction for any given input sample. Ideally, an attacker should achieve 100 \% fidelity, where the substitute and victim model agree on all the prediction output.
\paragraph{Accuracy Under Attack (\%)} It is defined as the percentage of adversarial test samples generated from substitute model being correctly classified by the victim model. It indicates the transferability of the adversarial examples as explained in prior~\cite{papernot2016transferability}. Ideally, if the substitute model and victim models are identical, then adversarial samples transferred from the substitute model should achieve similar efficacy (i.e., accuracy under attack ) as a white-box attack (i.e., the attacker knows everything about the victim model). In this evaluation, we use the popular projected gradient descent (PGD)~\cite{madry2018towards} attack to generate adversarial samples on the substitute model. The PGD attack uses $L_{\infty}$ norm, $\epsilon = 0.031$ and an attack iteration step of 7 for all three dataset.

\subsection{Hardware Configuration}

We train our DNN models using GeForce GTX 1080 Ti GPU platform operating at 1481MHz and deploy the trained models in an inference testbed. The \hleak attack is evaluated on the inference testbed equipped with Intel Haswell series processor (i5-4570) with AVX2 instruction set support.
We collect bitflip profile of the memory modules (i.e., templating) used in target system to identify potential vulnerable locations in DRAM. Note that memory templating is considered as a standard processor for rowhammer attack and we leverage existing techniques as described in~\cite{another_flip,eccploit,flip-fenshui,rambleed}.
The system is configured with {a dual channel memory subsystem with one 4GB DDR3 DIMM (Hynix) in each channel}. Our tested DIMMs have 71\% of the pages containing at least one $V_c$ and in total 0.017\% of memory cells are vulnerable to bit flip. Compared to bit flip profiles computed in prior work showing multiple DRAM modules with more than 98\% of all rows being vulnerable~\cite{hammertime}, our system has moderate level of vulnerability in rowhammer-induced bit flips.

\subsection{Dataset and Architecture}

We take three visual datasets: CIFAR-10~\cite{krizhevsky2010cifar}, CIFAR-100~\cite{krizhevsky2009learning} and German Traffic Sign Recognition Benchmark (GTSRB)~\cite{Stallkamp2012} for object classification task. CIFAR-10 contains 60K RGB images in size of $32\times32$. Following the standard practice, 50K examples are used for training and the remaining 10K for testing. On the other hand, CIFAR-100 has 100 classes with 600 images in each class. Both CIFAR-10 and CIFAR-100 has same image size and test-train data split. For GTSRB dataset, we use 40k labelled images and split them 85-15 ratio for training and evaluation purposes. Each traffic sign image has a size of 112x112 and is distributed in 42 different class. For CIFAR-10 experiments, we used residual networks (e.g., ResNet-18/34,Wide-ReNet-28)~\cite{he2016deep,zagoruyko2016wide} and VGG-11~\cite{simonyan2014very} architectures. For GTSRB and CIFAR-100, we adopted ResNet-18 and ResNext-50-32x4d~\cite{DBLP:journals/corr/XieGDTH16} as the evaluation architecture respectively. For all experiments, the weights of each victim model are quantized into 8 bit.

Following the standard practice in~\cite{cubuk2020randaugment,cubuk2019autoaugment}, we assume the attacker has access to a publicly available portion (i.e., $\sim$8\%) of the train dataset. In our experimental setting, we used 4096 ($\sim$8\%) train samples to train the substitute model for both CIFAR-10 \& CIFAR-100 dataset. For GTRSB, we only used 2656 ($\sim$8\%) train samples to train the substitute model based on our proposed \strain training method. We follow similar data distribution and experimental setting for the baseline method (i.e., architecture only case) as well. During the experiments, to train the substitute model for all different cases, we train three individual models independently and report the average (e.g., accuracy,fidelity) of three rounds.

\section{Evaluation}

\subsection{DNN Weight Recovery using \hleak}
\label{subsec:case_study}

As a practical example of \hleak, we choose an inference server as victim application running inference on different DNN models using \textbf{PyTorch} framework. Depending on the hardware instruction set supported on the host system, PyTorch performs extensive optimization of core computational kernel of DNN. For example, PyTorch has direct compatibility with FBGEMM~\cite{fbgemm} backend for x86 platform and QNNPACK~\cite{qnnpack} backend for ARM platform to accelerate matrix computation process using vector instructions (i.e., AVX-2 or AVX-512). Since there are multiple platform- and hardware-specific optimization for each specific type of DNN model following different execution path, for the rest of this discussion, we use 8-bit quantized DNN models running inference using FBGEMM backend as a case-study. We use PyTorch \texttt{v1.7.1-rc3} with FBGEMM commit \texttt{1d71039} running on Ubuntu 20.04 as the platform of our investigation.
Note that the similar inference execution flow can be obtained for other cases as well.

\begin{figure}[t]
    \centering
    \includegraphics[width=0.41\textwidth]{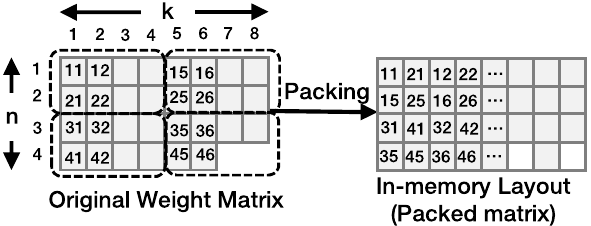} 
    \caption{Illustration of PyTorch weight packing (example using $4\times2$ size chunks) and organization of weights in memory. Each chunk is represented by dashed block and the small white blocks represents uninitialized location (no weight in this location).  }
    \label{fig:matrix_pack}
\end{figure}

\vspace{1mm}
\noindent\textbf{In-Memory Organization of DNN Weights.}
The accelerated vector instructions use single instruction multiple data (SIMD) for its internal calculation, this requires the operands used in these computation to be arranged in a specific layout (called \textit{packed} layout which is essentially reorganization of matrix data into a data array format optimized for sequential access by computational kernel~\cite{fbgemm}). PyTorch hence creates an optimized packed layout for both operands of these computation (i.e., weights and inputs). Since DNN model weights remain unchanged throughout an inference operation, PyTorch creates a packed copy for DNN weights of each linear and convolutional layers during model initialization and uses that pre-packed structure (stored in main memory) for each inference operation. Whenever a new inference is requested, PyTorch uses the pre-packed weights from main memory to carry out the inference. This pre-packing operation is instantiated by \texttt{PackedLinearWeight::prepack} and \texttt{PackedConvWeight::prepack} functions for linear and convolutional layer respectively and the packing operation is handled by \texttt{fbgemm::PackBMatrix} (FBGEMM uses \texttt{PackB} to represent weight matrices internally). FBGEMM divides the weights of each layer into smaller chunks, and then each chunk is stored sequentially into memory, which is useful for both accelerating computation performance and optimizing cache accesses. For AVX2 systems, each of these smaller chunks has a size of $512\times8$, indicating all weights in a specific layer are divided into several $512\times8$ weight chunks and weights in one chunk are laid sequentially into memory using \textit{column-major} format (data in columns are stored sequentially) forming a data array type layout which is different from regular memory layout of matrix that uses \emph{row-major} format~\cite{langsam1996data}. \Cref{fig:matrix_pack} illustrates the packing and memory layout of the stored weights. Given the a specific memory byte in the in-memory layout, the actual weight location in original weight matrix can be easily determined. Then within each layer, the matrix computational kernel performs vector multiplication and accumulation to produce the output of that layer.

\vspace{1mm}
\noindent\textbf{Mounting \hleak on PyTorch.}
We first use the \texttt{forward(<input>).toTensor()} method of DNN model as an anchor to monitor the beginning of an inference. This anchor will trigger the monitoring of subsequent anchors.
During the inference operation, PyTorch uses the \texttt{apply\_impl} function (corresponding to linear or convolutional layer) to instantiate the FBGEMM computation for each layer. We setup two anchor points (using FLUSH+RELOAD based monitoring) in each of the \texttt{apply\_impl}. Since the computation of each layer is sequential, the first call to \texttt{apply\_impl} represents the beginning of the first layer computation, while the second call denotes the second layer and so on. This is how we can distinguish between different layer computations and determine which layer is currently being executed. This passes a pointer to input and packed weights to the FBGEMM backend using \texttt{fbgemmPacked}. The \texttt{ExecuteKernel::execute} coordinates the generation of JIT code (\texttt{jit\_micro\_kernel\_fp}) for computation by executing the \texttt{getOrCreate} function. We setup another anchor point monitoring the access to \texttt{getOrCreate} function. We use batched victim page massaging by releasing vulnerable pages once this anchor is triggered to release pages of small number so that per-cpu pageset does not get overflown, but still have sufficient pages released for all the ML pages accessed in that chunk computation to occupy.

\subsection{\hleak Performance Analysis}

\begin{figure}[t]
	\centering
    \subfloat[b][Layer 1]{
	\includegraphics[width=0.45\textwidth]{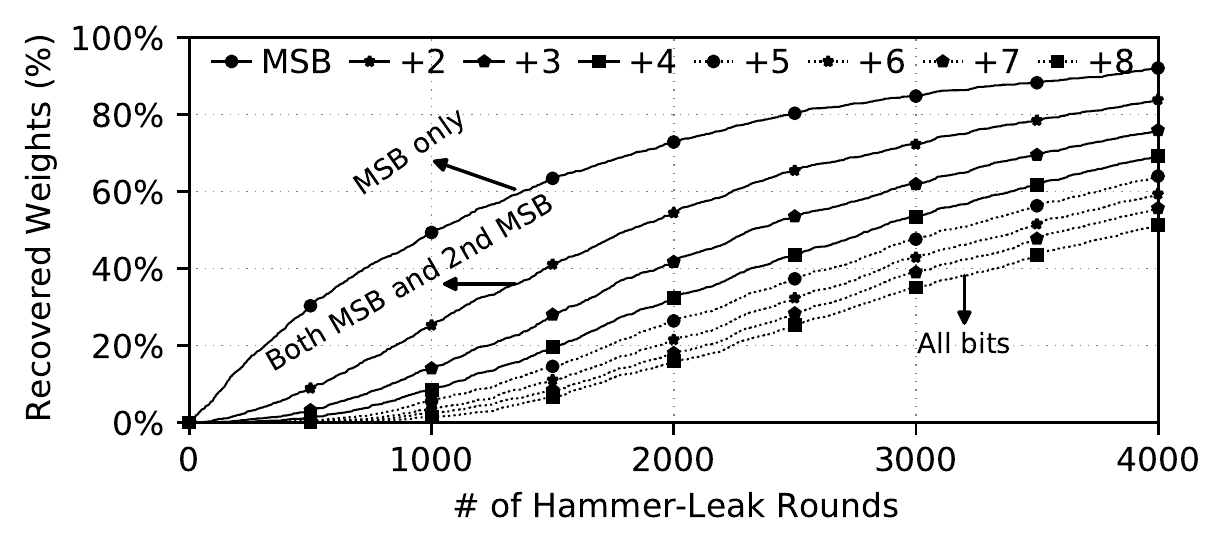}
	\label{fig:to_recov_1}}
\vspace{-1em}
	\subfloat[b][Picking a random layer (Layer 5 as an example)]{
	\includegraphics[width=0.45\textwidth]{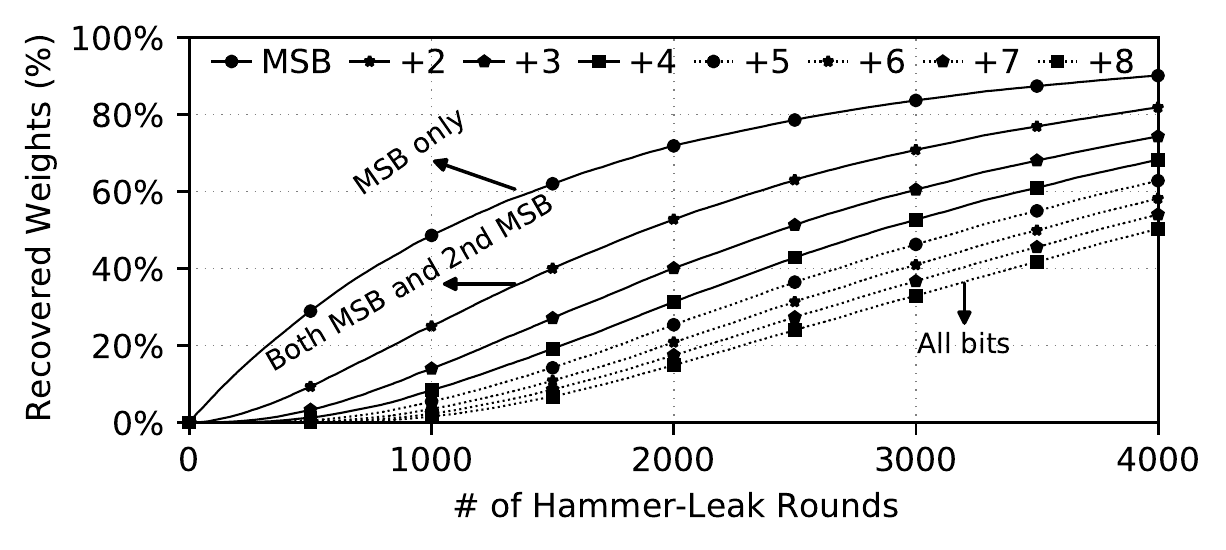}
	\label{fig:to_recov_5}}
	\caption{Percentage of weights with \texttt{MSB} and \texttt{MSB+x} bits recovered where \texttt{+x} represents \texttt{x} number of consecutive higher order bits recovery (i.e., \texttt{+3} represents weights with all three \texttt{MSB} bits recovered).}
	\label{fig:to_recov}
\end{figure}

In this section, we use {ResNet-18} as one representative DNN model for \hleak analysis. 
{ResNet-18} {has 21 layers with 11 million weight parameters}. We perform \hleak on this model to investigate the efficiency of our attack in recovering model parameters. We observe that at about 4000 \hleak rounds, \hleak can steal more than 90\% of the MSB bits for model weights almost across all layers (with the lowest per layer recovery rate to be 88\%). \Cref{fig:to_recov} shows the percentage of weights with leaked \texttt{\{MSB\}} bits as well as percentage of weights with other bits simultaneously leaked together with the MSB bits (e.g., \texttt{\{MSB+$2^{nd}$ MSB\}}) for two different layers. 
We observe that along with MSB bits, the recovery rate for additional weight bits are also very high, with 55\%-63\% weights across all layers have the complete weight recovered. This shows the high efficiency of the proposed targeted \hleak attack when deployed with the augmented rowhammer leakage.


\begin{figure}[t]
    \centering
    \includegraphics[width=0.43\textwidth]{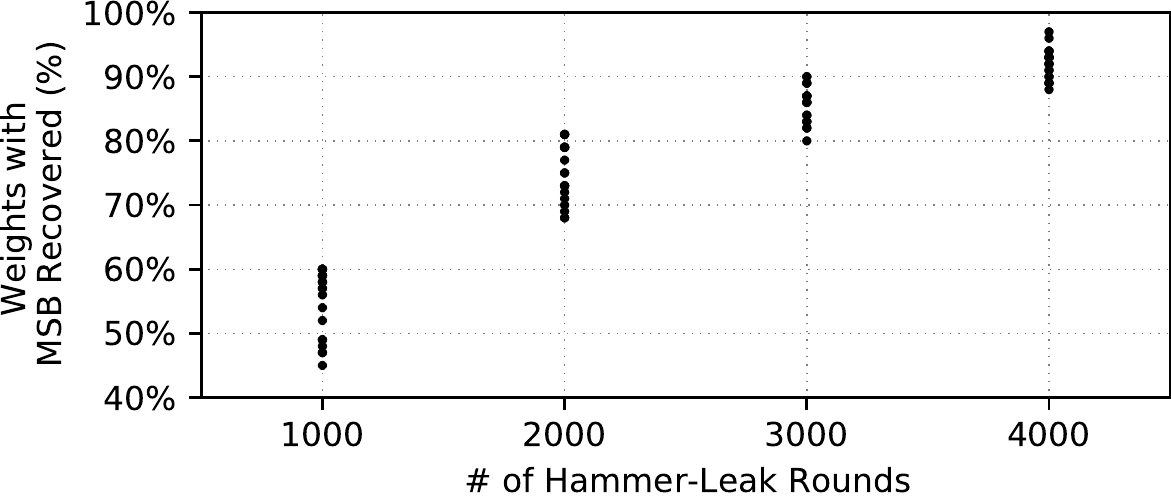} 
    \caption{Distribution of weights with \texttt{MSB} recovered across 21 individual layers of ResNet-18. }
    \label{fig:dist}
\end{figure}

\Cref{fig:dist} illustrates the distribution of weight percentages that has at least MSB bit recovered for all 21 layers. Depending on the attacker's goal (in our case, high percentage of weights with MSB exfilrated), \hleak can be completed much sooner than 4000 rounds. We observe most of the layers have half of the weights with MSB bit recovered within 1000 rounds of attack.

We further investigate the time spent during each of the steps of a \hleak round to determine the attack cost factors. We experimentally find that memory exhaustion (Step 1) and Bitflip-aware page release (Step 2) requires 12 seconds and 21 seconds respectively. The actual inference operation (which is Step 3 when the inference application accesses the ML pages) takes less than 1 seconds. Finally, by far the most expensive portion (in terms of latency overhead) is the actual bit leakage step (Step 4) where \hleak steals the model bits through rowhammer-based side channel. When considering bit leakage from pages that have at least one MSB bit offset in $V_c$ (i.e., \textit{MSB} configuration in \Cref{tab:cifar10_res}), each \hleak round will hammer about 11K rows on average, which takes on average aout 239 seconds. In contrast, without constraining MSB pages, recovering all bits (i.e., the \textit{All Bits} configuration in \Cref{tab:cifar10_res}) requires hammering 17K rows on average (i.e., about 375 seconds). Note that only Step 2 and Step 3 have to be done simultaneously with the inference operation, the bulk of the operation (Step 4) can be done in-between multiple inferences by limiting the number of vulnerable pages to release in each round. In addition, instead of considering each vulnerable page while releasing, we can only release the physical pages that have at least one MSB offset leakable (i.e., MSB prioritization scheme). This way we can substantially improve the attack efficiency while leaking most of the influential bits. We experimentally found that using MSB prioritization scheme, \hleak can still recover 68\% other bits along with recovering 92\% of MSB bits. 

\begin{table*}[ht]
\centering
\caption{\emph{Summary of CIFAR-10 results for three different DNN architectures. We report two different cases of \dsteal attack i) All Bits: where we use all the bit information (i.e., all 8 plots) plotted in \Cref{fig:to_recov}. According to this plot, for each \# of \hleak attack rounds along x-axis, we take the percentage of bits recovered for all 8 plots (e.g., MSB,MSB+$2^{nd}$ MSB \& so on). ii) MSB: We only use the MSB bit information labelled as MSB curve in \Cref{fig:to_recov}. }}
\label{tab:cifar10_res}
\scalebox{0.89}{
\begin{tabular}{@{}ccccccccccccc@{}}
\toprule
\multicolumn{4}{c}{\emph{Method}} & \multicolumn{3}{c}{\emph{ResNet-18}} & \multicolumn{3}{c}{\emph{ResNet-34}} & \multicolumn{3}{c}{\emph{VGG-11}} \\ \midrule
\begin{tabular}[c]{@{}c@{}}\# of Hammer-Leak\\ Rounds \end{tabular} & Method & Case  &
 \begin{tabular}[c]{@{}c@{}}Time\\ (days)\end{tabular} &
\begin{tabular}[c]{@{}c@{}}Accuracy\\ (\%)\end{tabular} & \begin{tabular}[c]{@{}c@{}}Fidelity\\ (\%)\end{tabular} & \begin{tabular}[c]{@{}c@{}}Accuracy\\ under\\ Attack\\ (\%)\end{tabular} & \begin{tabular}[c]{@{}c@{}}Accuracy\\ (\%)\end{tabular} & \begin{tabular}[c]{@{}c@{}}Fidelity\\ (\%)\end{tabular} & \begin{tabular}[c]{@{}c@{}}Accuracy\\ Under\\ Attack\\ (\%)\end{tabular} & \begin{tabular}[c]{@{}c@{}}Accuracy\\ (\%)\end{tabular} & \begin{tabular}[c]{@{}c@{}}Fidelity\\ (\%)\end{tabular} & \begin{tabular}[c]{@{}c@{}}Accuracy\\ Under\\ Attack\\ (\%)\end{tabular} \\ \midrule
Baseline & Architecture Only &-&-&  72.68& 73.59 & 62.58 & 72.59 & 73.24 & 63.53 & 70.9 & 71.52 & 62.08 \\ \midrule
\multirow{2}{*}{1500} & \multirow{2}{*}{\dsteal} &All Bits& 7.5 & 75.9 & 77.13 & 54.49 & 75.08 & 75.65 & 56.23 & 72.35 & 73.24 & 63.53 \\& &MSB& 3.9 & 77.0 & 78.24 & 54.8 & 76.97 & 77.89 & 55.14 &  72.81 & 74.04 & 58.27\\ \midrule
\multirow{2}{*}{3000} & \multirow{2}{*}{\dsteal} &All Bits & 14.9 & 86.28 & 87.54 & 5.69 & 85.15 & 86.08 & 5.28 & 81.57 & 82.74 & 35.85 \\
 & & MSB & 7.8 & 85.39 & 86.97 & 12.49 & 87.03 & 88.2 & 7.61 & 77.24 & 78.48  & 37.6 \\\midrule
\multirow{2}{*}{4000} & \multirow{2}{*}{\textbf{\emph{\dsteal}}}  &All Bits& \textbf{\emph{17.3}} & \textbf{\emph{89.15}} & \textbf{\emph{90.64}} & \textbf{\emph{1.89}} & \textbf{\emph{88.02}} & \textbf{\emph{89.22}} & \textbf{\emph{1.69}} & \textbf{\emph{84.67}} & \textbf{\emph{86.55}} & \textbf{\emph{17.2}} \\
 &   &MSB& \textbf{\emph{10.4}} & \textbf{\emph{88.65}} & \textbf{\emph{90.6}} & \textbf{\emph{2.21}} & \textbf{\emph{89.42}} & \textbf{\emph{90.58}} & \textbf{\emph{2.12}} & \textbf{\emph{79.84}} & \textbf{\emph{80.42}} & \textbf{\emph{26.44}} \\\midrule
Best-Case &  White-box &-&- & 93.16 & 100.0 & 0.0 & 93.11 & 100.0 & 0.0 & 89.96 & 100.0 & 4.63 \\ \bottomrule
\end{tabular}}
\end{table*}

\subsection{\dsteal Experimental Results: CIFAR-10}
In \Cref{tab:cifar10_res}, we evaluate the performance of \dsteal attack on CIFAR-10 dataset for three different architectures. Further, we show an ablation study showing the impact of using several rounds of \hleak attack information for \dsteal attack. As a baseline method, we only compare with the architecture only case (i.e., 0-bit information leaked). For the baseline case, we assume the attacker only knows the victim model architecture. Then, a substitute model with the same architecture is trained using a similar setting (i.e., less than \emph{8} \% available data). On the other hand, we treat the white-box case as the best-case scenario where the attacker knows every information (i.e., weights, biases and architecture) of the victim model. In summary, 
with more and more recovered weight bits, our \dsteal achieves much improved accuracy, fidelity and adversarial example attack efficacy. For example, our substitute model can generate effective transferable adversarial example with similar efficacy (i.e., $\sim$\emph{0} \%) as white-box attack for both ResNet-18 and ResNet-34.

In our evaluation, the residual victim models ResNet-18 \& ResNet-34 have 93.16 \% \& 93.11 \% inference accuracy, respectively. As shown in \Cref{fig:to_recov}, after 4000 rounds of \hleak attack, the attacker could recover 90 \% of the MSB bits ($\sim$11.52 \% of total bits). Next, by only using the leaked MSB bit information, the attacker can recover up to 88.65/89.42 \% test accuracy for the ResNet (18/34) models. However, for All Bits case in \Cref{tab:cifar10_res}, even after paying an additional time cost (i.e., \emph{1.66} $\times$), the performance of \dsteal attack can hardly improve on residual models. In contrast, a larger (i.e., 132 Million parameters) model, such as VGG-11 with different architecture topology, highly benefits from the additional information of all the bits. For VGG, we observe a $\sim$\emph{5} \% improvement by using all the filtered bits in comparison to using MSB only. Similar to the 4000 round case, we observe a similar pattern in accuracy recovery at 3000 rounds of \hleak attack as well.

Next, we evaluate the adversarial attack performance for both 3000 and 4000 rounds of \hleak attack. In \Cref{tab:cifar10_res}, we show that our substitute model can transfer effective adversarial samples to the victim model across all three architectures. In particular, for ResNet models (e.g., 18/34), our substitute model generated adversarial examples demonstrate close to white-box attack efficacy (i.e., within \emph{2-5} \% of the best case result). As for VGG model, which is already known as a robust architecture~\cite{madry2018towards}, our substitute model generated adversary reaches within $\sim$\emph{13-22} \% of an ideal white-box attack. Nevertheless, our attack efficacy still shows an improvement of about $\sim$\emph{45-60} \% across all three architectures in comparison to the baseline (i.e., architecture only) technique. 

Finally, we consider an attack scenario where the attacker has a strict time budget. In this scenario, let's assume he/she can only afford to run 1500 rounds of \hleak attack while prioritizing MSBs (e.g., only \emph{3.9} days of attack time). As summarized in \Cref{tab:cifar10_res}, even such a restricted attack can generate effective adversarial examples to lower accuracy under attack by \emph{8\%} for ResNet models and by \emph{4\%} for VGG-11 compared to baseline. One key observation for this low budget (i.e., 1500 round attack) attack is that attacker can generate a much more effective substitute model by only using MSB information rather than all the bits. The reason being with limited bit information (e.g., 50 \%  MSB only) putting strict penalization (i.e., mean clustering) on the weights during training does not help the substitute model accuracy. In fact, for VGG-11, it becomes worse than the baseline method. As a result, for \dsteal attack with limited partial bit information, using the relaxation of the weight constraints (i.e., MSB only) can be more effective than using all the available filtered bits.

\begin{table*}[ht]
\centering
\caption{\emph{Summary of \dsteal attack performance on large-scale dataset (i.e., CIFAR-100: 100 output class \& GTSRB: 112x112 input dimension). Our proposed \dsteal outperforms the baseline (i.e., architecture only) case across all three evaluation metrics even on large-scale image dataset. The improvements are shown in comparison to the baseline method.}}
\label{tab:other_res}
\begin{tabular}{@{}ccccccc@{}}
\toprule
\emph{Dataset} & \multicolumn{3}{c}{\emph{CIFAR-100}} & \multicolumn{3}{c}{\emph{GTSRB}} \\ \midrule
 Method & \begin{tabular}[c]{@{}c@{}}Accuracy\\ (\%)\end{tabular} & \begin{tabular}[c]{@{}c@{}}Fidelity\\ (\%)\end{tabular} & \begin{tabular}[c]{@{}c@{}}Accuracy\\ under\\ Attack\\ (\%)\end{tabular} & \begin{tabular}[c]{@{}c@{}}Accuracy\\ (\%)\end{tabular} & \begin{tabular}[c]{@{}c@{}}Fidelity\\ (\%)\end{tabular} & \begin{tabular}[c]{@{}c@{}}Accuracy\\ Under\\ Attack\\ (\%)\end{tabular} \\ \midrule
Baseline & 32.81 & 33.97 & 42.7 & 98.67 & 98.01 & 68.28 \\
\textbf{\emph{\dsteal}} & \textbf{\emph{59.8 ($\uparrow$ 26)}} & \textbf{\emph{64.11($\uparrow$ 31)}} & \textbf{\emph{6.61}} & \textbf{\emph{99.57}} & \textbf{\emph{98.77}} & \textbf{\emph{43.5($\downarrow$ 24)}} \\
White-Box & 69.7 & 100.0 & 0.0 & 99.05 & 100.0 & 4.32 \\ \bottomrule
\end{tabular}
\end{table*}

\subsection{\dsteal Experimental Results: CIFAR-100 \& GTSRB}
We evaluate \dsteal attack on two relatively larger datasets as well. We chose CIFAR-100, which has \emph{10$\times$} larger output class size than CIFAR-10, and the GTSRB dataset has \emph{12$\times$} larger input dimension than CIFAR-10. In \Cref{tab:other_res}, we summarize the results of these two datasets. 

For CIFAR-100, we attack a ResNext-50-32x4d victim DNN model. Here, the baseline (i.e., architecture only) method can recover the model accuracy only up to \emph{32.81} \% which is $\sim$\emph{36} \% lower than the ideal case (i.e., white-box). Our attack can improve this baseline recovered accuracy by $\sim$\emph{26} \% and also the fidelity by $\sim$\emph{31} \%. Again, for GTSRB, our attack can recover the exact baseline accuracy with a high fidelity rate \emph{98.77 \%}. It shows the dataset with a larger output class (e.g., CIFAR-100) poses a tough challenge to recover the test accuracy of the victim model. 

As for adversarial input attack, our substitute model can achieve \emph{6.6} \% accuracy under attack on CIFAR-100. It shows almost similar attack efficacy (i.e., \emph{0-6} \%) as a white-box attack while gaining a \emph{36} \% improvement compared to baseline. For GTSRB, our substitute model can lower the accuracy under attack by $\sim$\emph{24} \% compared to the baseline. However, the attack efficacy still lags by \emph{39} \% from the ideal case (i.e., white-box) mainly because of the input image dimension (112$\times$112). As an adversarial image generated by the substitute model is less likely to succeed in attacking the victim model for a larger input dimension/search space.

\begin{table*}[ht]
\centering
\caption{\emph{We evaluate \dsteal attack against state-of-the-art techniques across three different domains as case studies. In each of the cases, only our attack performs on-par with the SOTA methods across all three evaluation metrics.}}
\label{tab:sota}
\scalebox{0.99}{
\begin{tabular}{@{}ccccccc@{}}
\toprule
\emph{Case Study}&\emph{Method} & \emph{Objective} & \emph{Model} & \begin{tabular}[c]{@{}c@{}}\emph{Accuracy}\\ \emph{(\%)}\end{tabular} & \begin{tabular}[c]{@{}c@{}}\emph{Fidelity}\\ \emph{(\%)}\end{tabular} & \begin{tabular}[c]{@{}c@{}}\emph{Accuracy}\\ \emph{Under}\\ \emph{Attack (\%)}\end{tabular} \\ \midrule
\multirow{4}{*}{\emph{Regularization}}
&Fully-Supervised~\cite{tramer2016stealing}& Accuracy/Fidelity & WideResNet-28 & 86.51 & 87.37  & - \\
\multirow{4}{*}{\emph{\& Data Augmentation}}&Rand-Augment~\cite{cubuk2020randaugment}& Accuracy & WideResNet-28 & 87.4 & - & - \\
&Auto-Augment~\cite{cubuk2019autoaugment}& Accuracy & WideResNet-8 & 87.7 & - & - \\
 &Mix-Match~\cite{jagielski2020high} & Accuracy/Fidelity & WideResNet-28 & 93.2 & 93.9  & - \\

&\textbf{\emph{\dsteal (ours)}} & \textbf{\emph{Accuracy/Fidelity/Attack}} & \textbf{\emph{WideResNet-28}} & \textbf{\emph{90.35}} & \textbf{\emph{91.82}} & \textbf{\emph{0.16}} \\ \midrule
\multirow{2}{*}{\emph{Model Extraction}}&Side Channel~\cite{hu2020deepsniffer,wei2020leaky,xiang2020open,244042}& Accuracy/Fidelity/Attack & ResNet-18 & 72.68 & 73.59 & 62.58 \\
&\textbf{\emph{\dsteal (ours)}} & \textbf{\emph{Accuracy/Fidelity/Attack}} & \textbf{\emph{ResNet-18}} & \textbf{\emph{90.02}} & \textbf{\emph{91.67}} & \textbf{\emph{1.2}} \\ \midrule
\multirow{3}{*}{\emph{Input Attack}}

&Black-Box (Inception-V1)~\cite{cui2020substitute}& Adversarial Attack & ResNet-18 & - & - & 20.47 \\
& White-Box (PGD/Trades)~\cite{madry2018towards,zhang2019theoretically}& Adversarial Attack & ResNet-18 & - & - & 0.0 \\ 

&\textbf{\emph{\dsteal (ours)}} & \textbf{\emph{Adversarial Attack}} & \textbf{\emph{ResNet-18}} & \textbf{\emph{-}} & \textbf{\emph{-}} & \textbf{\emph{1.2}} \\
\bottomrule
\end{tabular}}
\end{table*}

\subsection{Comparison to State-of-the-art Techniques}
In \Cref{tab:sota}, we summarize the standing of our \dsteal attack compared with existing model recovery methods for three different domains of applications. First, existing model regularization~\cite{tramer2016stealing,jagielski2020high} and data augmentation techniques~\cite{cubuk2020randaugment,cubuk2019autoaugment} are useful in training deep models with limited data. Compared to them, our substitute model achieves a much higher accuracy(i.e., $\sim$\emph{3}\%). The only exception is Mix-Match~\cite{jagielski2020high}, as it performs slightly better (i.e., $\sim$\emph{2}\%) than our proposed technique. However, Mix-Match has adopted a combination of existing data augmentation and regularization techniques, which could also be integrated with our method to boost the performance. Hence, neither regularization nor data augmentation methods is a competing method to our attack. In addition, the mix-match recovery method requires significant modification to make it functionally equivalent only for a two-layer neural network, as discussed in~\cite{jagielski2020high}.

Other existing side channel attacks~\cite{hu2020deepsniffer,wei2020leaky,xiang2020open,244042,yu2020deepem} fall into a similar attack surface category as \dsteal. Among them,~\cite{yu2020deepem} is only applicable to binary neural networks. On the contrary, our attack is a more general version of the attack applicable to any bit-width. Other side channel attacks~\cite{hu2020deepsniffer,wei2020leaky,xiang2020open,244042} focus on recovering the architecture and then training the model with limited data. To compare with them, we assume the attacker knows the exact model architecture, which is on top of those attacks. Then, our \dsteal leverages the leaked weight bits to further improve the attack efficacy. From this point, our attack certainly outperforms such prior architecture only model extraction attacks with $\sim$18 \% improvement in accuracy and $\sim$61 \% improvement in degrading the accuracy under adversarial attack.

Finally, \dsteal attack falls into a grey zone between black-box and white-box adversarial input attack. In \Cref{tab:sota}, we show \dsteal can achieve 1.2 \% accuracy under attack, which is more closer to a white-box attack performance (i.,e., 0 \%). We observe a 17 \% improvement in attack performance compared to a powerful black-box substitute model (e.g., Inception-V1) attack.

\section{Countermeasures for \dsteal}

\subsection{Hardware-Assisted Protection for Secret Pages} 
One possible way to mitigate \dsteal information leakage is to protect the secretive pages through support of trusted execution environments such as Intel SGX~\cite{costan2016intel}. With Intel SGX, pages that belongs to a protected enclave is encrypted using processor-side memory encryption engine (MEE) before leaving the processor die. This means that the enclave pages stored in the main memory are encrypted by design and hence using this attack, the attacker cannot retrieve any information about the actual data. However, applications running on enclave have several times of overhead in runtime~\cite{orenbach2017eleos} compared to unsecure non-enclave version because of the nature of protection. Additionally,
the maximum effective size of encrypted data size in memory is 96MB in Microsoft Windows~\cite{epc}, which creates non-trivial overhead for applications requiring larger memory than this (i.e., large DNN models), due to the necessarily complex mechanism of page-fault handling. 

\subsection{Software-based Protection of DNN Weights}
Alternative to the hardware-assisted protection schemes, application developer can use software based encryption-decryption scheme for streaming applications. For example, using software base en-/decryption, the ML framework can perform on-demand weight decryption during computation and does not store the decrypted weight at any time. Since the unencrypted weights are not stored in memory, \hleak will unable to any secret on this. Note that while the runtime overhead of such software-based protection scheme needs evaluation, one block encryption (128-bit) using AES-NI instruction set (advanced instruction set designed specifically to accelerate AES encryption/decryption) still requires 8-cycle latency~\cite{aesni}. This can accumulate substantially and result in potentially high run-time overhead, limiting the deployment of such implementation in DNN applications.


\section{Conclusion}
The training of deep neural networks requires heavy computational resources and sensitive private data. Thus, any potential breach in model privacy through leakage of sensitive model parameters may cost the service provided a heavy penalty. Our proposed \dsteal attack reveals the serious threat of an effective model extraction attack. In particular, our weight bit extraction method \hleak showed it could recover a significant portion of the weight bits. Our substitute model training algorithm can leverage this information to effectively launch a strong adversarial input attack to the victim model. The success of \dsteal attack demands more defense work in protecting the IP of DNN models. 

\bibliographystyle{IEEEtran}
\bibliography{IEEEexample,ref_reyad}

\end{document}